\documentclass[preprint,aps,superscriptaddress,floatfix,longbibliography]{revtex4-1}
\usepackage{graphicx,epsfig,subfig,dcolumn,bm,mathrsfs,amsmath,amsthm,color}
\usepackage{hyperref}

\newcommand{\change}[1]{\textcolor{black}{#1}}

\newcommand{\ud}{\mathrm{d}}

\begin{document}
\title{Wetting dynamics in an angular channel}

\author{Chen Zhao}
\affiliation{Center of Soft Matter Physics and its Applications, Beihang University, Beijing 100191, China}
\affiliation{School of Physics, Beihang University, Beijing 100191, China}

\author{Tian Yu}
\affiliation{State Key Laboratory of Hydroscience and Engineering, Tsinghua University,
Beijing 100084, China}

\author{Jiajia Zhou}
\email[]{zhouj2@scut.edu.cn}
\affiliation{South China Advanced Institute for Soft Matter Science and Technology, South China University of Technology, Guangzhou 510640, China}
\affiliation{School of Molecular Science and Engineering, Guangzhou International Campus, South China University of Technology, Guangzhou 511442, China}

\author{Masao Doi}
\affiliation{Center of Soft Matter Physics and its Applications, Beihang University, Beijing 100191, China}

\date{\today}

\begin{abstract}
We analyze the dynamics of liquid filling in a thin, slightly inflated rectangular channel driven by capillary forces.
We show that although the amount of liquid $m$ in the channel increases in time following the classical Lucas-Washburn law, $m \propto t^{1/2}$, the prefactor is very sensitive to the deformation of the channel because the filling takes place by the growth of two parts, the bulk part (where the cross-section is completely filled by the liquid), and the finger part (where the cross-section is partially filled).
We calculate the time dependence of $m$ accounting for the coupling between the two parts
and show that the prefactor for the filling can be reduced significantly by a slight deformation of the rectangular channel, e.g., the prefactor is reduced 50\% for a strain of 0.1 \%.
This offers an explanation for the large deviation in the value of the prefactor reported previously.
\end{abstract}

\maketitle

\section{Introduction}

Spontaneous filling of water into a glass tube is the hallmark of the capillarity and wetting phenomena \cite{dBQ, deGennes1985, Bonn2009}.
The dynamical theory of the capillary filling was given by Lucas \cite{Lucas1918} and Washburn \cite{Washburn1921} more than one century ago: the balance between the capillary force and the viscous friction leads to the well-known $t^{1/2}$ scaling of the filling length
\begin{equation}
  h(t) = \mathscr{C}_{\rm LW} \, t^{1/2}.
\end{equation}
For a circular tube with radius $r$, the prefactor is given by $\mathscr{C}^2_{\rm LW} = r \gamma \cos\theta / (2\eta) $; for a rectangular channel of small aspect ratio (short side length $b \ll$ long side length $2a$), the prefactor becomes $\mathscr{C}^2_{\rm LW} = b \gamma \cos\theta /(3\eta)$; where $\gamma$ is the surface tension of the fluid, $\theta$ is the equilibrium contact angle (henceforth assumed to be zero), and $\eta$ is the fluid viscosity.

The Lucas-Washburn law is well established in the macroscopic scale (characteristic length greater than micrometer).
With the advance in microfabrication techniques, recent studies started to explore the region of nanometer scale.
Such studies show unanimously that the $t^{1/2}$ scaling remains valid in the nanoscale, but the prefactor varies significantly (by factor 2 or 3) depending on the system
\cite{Tas2004, vanDelft2007, Persson2007, Haneveld2008, Hamblin2011, Chauvet2012}.
Most experiments observed a reduction of the prefactor with respect to the Lucas-Washburn prediction.
Various possible explanations have been proposed, including the dynamic contact angle, precursor film, trapped gas bubbles, etc. (see Ref. \cite{Chauvet2012} for a review).

A glance over recent literature reveals that most experiments were conducted in a rectangular channel with corners.
It is known that in the cornered geometry, finger-like part develops along the corners
and advances ahead of the bulk part \cite{Concus1969, Weislogel2012, 2021_rect_sstar}.
The transition region between the bulk and the finger is of finite size (on the order of the
channel width) and can be ignored at a late stage where the lengths of the bulk and the finger become much larger than the channel width.
In such a case, one can assume that the finger part starts with a specific saturation $s^*$,
defined as the fraction of the cross-section area occupied by the fluid \cite{2018_square, 2021_rect_sstar}.

Another interesting observation is that the rectangular channels used in these experiments have small aspect ratio, with a short side $b$ in the range of a few nanometers and a long side $2a$ of size a few micrometers, leading to an aspect ratio $\beta=b/a$ in the range of 0.001--0.1.

In the previous work \cite{2021_rect_sstar}, we have shown that the saturation $s^*$ of rectangular channels of small aspect ratio is very sensitive to the shape change of the cross section.
For example,  $s^*$ changes from $0.001$ to $0.7$ when the long sides of the rectangle become non-parallel, making an angle of 0.01 radian for the channel of aspect ration $b/a=0.02$.
In this paper, we shall show that this change in $s^*$ has a significant effect on the dynamics of capillary filling.
We shall investigate the coupled dynamics of the bulk and the finger, and calculate the prefactor in the Lucas-Washburn equation precisely.
We shall show that the prefactor is very sensitive to the paralleling of the opposite sides of the rectangular channel.
This offers a plausible explanation for the reduction of the prefactor reported in many previous experiments.

\section{Methods}

\begin{figure}[htp]
\includegraphics[width=0.8\columnwidth]{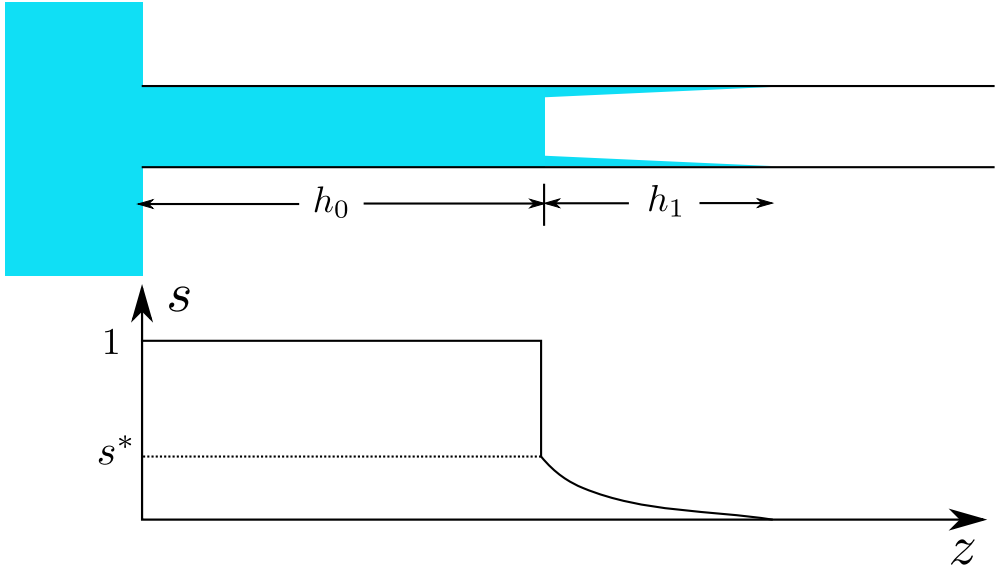}
\caption{Capillary filling in a channel with corners. The bulk part has a full saturation $s=1$ and its length is $h_0$. The finger part starts with a partial saturation $s^*$ and the finger length is $h_1$.}
\label{fig:sketch}
\end{figure}

First we formulate the dynamics of liquid flow in a straight channel with
corners.
We take $z$ coordinate along the channel, and define the saturation $s(z)$ as the ratio of the
area occupied by the liquid to the total area in the cross section at $z$.
Figure~\ref{fig:sketch} shows the schematic picture of $s(z;t)$.
The saturation $s(z;t)$ is equal to 1 in the bulk part ($0<z<h_0(t)$), and is less than 1 in the finger part ($h_0(t)<z<h_0(t)+h_1(t) $).
Since the transition from the bulk to finger takes place in a region which is much smaller than the lengths of bulk and finger, it is ignored in Fig.~\ref{fig:sketch}.
As it is shown in Fig.~\ref{fig:sketch},  the finger starts at $z=h_0(t)$ taking the value of $s^*$, i.e., $s(h_0(t), t)= s^*$.

Such assumption can be justified generally for any cornered channels with the coexistence of the bulk and the finger, as long as the filling length is much greater than the channel size in the cross-section so the transition region between the bulk and the finger is of finite size.

We shall derive coupled time-evolution equations for $h_0(t)$ and $s(z;t)$ using Onsager variational principle \cite{DoiSoft}.

The free energy of the system is given by
\begin{equation}
  A = h_0 f(1) + \int_{h_0}^{h_0+h_1} f(s) \ud z \, ,
\end{equation}
where $f(s)$ is the interfacial free energy density (per unit length) that is a function of the local saturation $s$.
The first term represents the free energy of the bulk part, and the second term represents that of the finger part.
The change rate of the free energy is
\begin{eqnarray}
  && \dot{A} = \dot{h}_0 f(1) - \dot{h}_0 f(s^*) + \int_{h_0}^{h_0+h_1} f'(s) \dot{s} \ud z \\
  && = \dot{h}_0 f(1) - \dot{h}_0 f(s^*) + f'(s^*) j_1^* + \int_{h_0}^{h_0+h_1} f''(s) j_1 \frac{\partial s}{\partial z} \ud z \, , \nonumber
  \label{eq:Adot}
\end{eqnarray}
where $f'(s) \equiv \ud f/ \ud s$ and $f''(s) \equiv \ud^2 f / \ud s^2$.
Here we have used the conservation of the fluid volume
\begin{equation}
  \label{eq:cons}
  \dot{s} = - \frac{\partial j_1}{\partial z} \, ,
\end{equation}
where $j_1$ is the volume flux in the finger divided by the cross-section area of the channel,
and $j_1^* \equiv j_1(h_0)$ is defined at the entrance of the finger.

The dissipation function also includes two terms
\begin{equation}
  \Phi = \frac{1}{2} \zeta(1) h_0 j_0^2 + \frac{1}{2} \int_{h_0}^{h_0+h_1} \zeta(s) j_1^2 \,\ud z \, ,
\end{equation}
where $j_0 = (1-s^*)\dot{h}_0 + j_1^*$ is the flux in the bulk and $\zeta(s)$ is the friction coefficient that depends on the local saturation.

The Rayleighian is given by $\mathscr{R} = \dot{A} + \Phi$
\begin{eqnarray}
  && \mathscr{R} = \dot{h}_0 f(1) - \dot{h}_0 f(s^*) + f'(s^*) j_1^* + \int_{h_0}^{h_0+h_1} f''(s) j_1 \frac{\partial s}{\partial z} \, \ud z\nonumber \\
  &&+  \frac{1}{2} \zeta(1) h_0 \left[ j_1^* + \dot{h}_0 (1-s^*) \right]^2 + \frac{1}{2} \int_{h_0}^{h_0+h_1}  \zeta(s) j_1^2 \ud z \, .
\end{eqnarray}
We now apply the Onsager variational principle to obtain the time-evolution equations for the bulk $h_0(t)$ and the finger $h_1(t)$.

The variation of $\mathscr{R}$ with respect to $j_1$ is
\begin{equation}
  \label{eq:dRdj1}
  \frac{\delta \mathscr{R}}{\delta j_1} = 0 \quad
  \Rightarrow \quad
  j_1 = - \frac{f''(s)}{\zeta(s)} \frac{\partial s}{\partial z} \, .
\end{equation}
\change{Combining} with the conservation equation (\ref{eq:cons}), we obtain a diffusion equation
\begin{equation}
  \label{eq:diffusion0}
  \dot{s} = \frac{\partial}{\partial z} \left[ D(s) \frac{\partial s}{\partial z} \right],
\end{equation}
with the diffusion constant $D(s) \equiv {f''(s)}/{\zeta(s)}$.
We perform a change of variable $z' = z-h_0$ and rewrite Eq.~(\ref{eq:diffusion0})
\begin{equation}
  \label{eq:diffusion}
  \frac{\partial s}{\partial t} = \frac{\partial}{\partial z'} \left[ D(s) \frac{\partial s}{\partial z'} \right] + \dot{h}_0 \frac{\partial s}{\partial z'}.
\end{equation}
This is a partial differential equation (PDE) of $s(z';t)$.
The boundary conditions are $s(z'=0) = s^*$ and $s(z'=h_1) = 0$.

The variation of $\mathscr{R}$ with respect to $j_1^*$ and $\dot{h}_0$ \change{leads} to
\begin{eqnarray}
  \label{eq:dRdj1star}
  & f'(s^*) + \zeta(1) h_0 \left[ j_1^* + (1-s^*) \dot{h}_0 \right] = 0 \, ,  & \\
  \label{eq:dRdh0}
  &f(1)-f(s^*) + \zeta(1) h_0 (1-s^*) \left[ j_1^* + (1-s^*) \dot{h}_0 \right] = 0 \, . &
\end{eqnarray}
From the above two equations we obtain
\begin{equation}
  \label{eq:star}
  f(1)-f(s^*) = (1-s^*) f'(s^*).
\end{equation}
This is just the definition of $s^*$ \cite{2021_rect_sstar}.

The flux at the finger entrance $z'=0$ is
\begin{equation}
  \label{eq:j1star}
  j_1^* = -D(s^*) \frac{\partial s}{\partial z'} \Big|_{z'=0}.
\end{equation}
Substituting the above expression into Eq.~(\ref{eq:dRdj1star}),
we arrive at an ordinary differential equation (ODE) for $h_0$
\begin{equation}
  \label{eq:jstar}
  \dot{h}_0 = \frac{1}{(1-s^*)} \left[ D(s^*) \frac{\partial s}{\partial z'} \Big|_{z'=0} - \frac{ f'(s^*) }{\zeta(1)} \frac{1}{h_0} \right] .
\end{equation}

The time-evolution of the system is governed by coupled PDE (\ref{eq:diffusion}) and ODE (\ref{eq:jstar}).
The dynamics of the finger profile $s(z';t)$ is given by the PDE (\ref{eq:diffusion}), but a closed solution cannot be obtained without knowing the bulk velocity $\dot{h}_0$.
The dynamics of the bulk length $h_0(t)$ is determined by the ODE (\ref{eq:jstar}), and again it depends on the slope of the saturation $(\partial s / \partial z')|_{z'=0}$ at the finger entrance.
\change{The procedure prensented here can be applied to channels with a general cross-sectional shape, as long as the free energy function $f(s)$ and the friction coefficient $\zeta(s)$ are known. 
The detailed calculation of the these two functions depend on the specific geometry of the channel (see supporting information for details and also Refs. \cite{2018_square, 2021_rect_sstar}.)}


If we ignore the finger flow, the Rayleighian is solely determined by the bulk flow
\begin{equation}
  \mathscr{R} = \dot{h}_0 f(1) + \frac{1}{2} \zeta(1) h_0 \dot{h}_0^2 \, .
\end{equation}
The time-evolution equation is given by $\delta \mathscr{R} / \delta \dot{h}_0=0$, which leads to
\begin{equation}
  h_0 \dot{h}_0 = - \frac{f(1)}{\zeta(1)} \, .
\end{equation}
The solution to the above equation with the initial condition $h_0(t=0)=0$ is the well-known Lucas-Washburn law
$h_0 =  \mathscr{C}_{\rm LW} t^{1/2}$ with the prefactor given by
\begin{equation}
  \label{eq:CLW}
  \mathscr{C}^2_{\rm LW} = \frac{ 2 |f(1)|} {\zeta(1)} \, .
\end{equation}

\section{Results and Discussion}

\begin{figure}[htp]
\includegraphics[width=0.7\columnwidth]{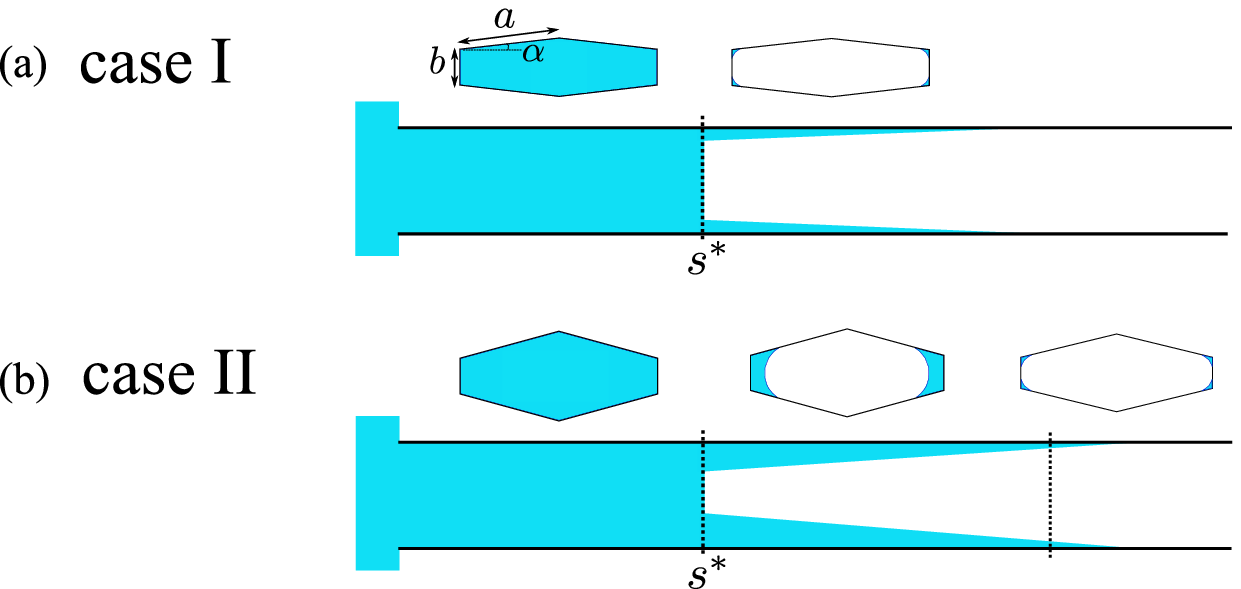}\\
\includegraphics[width=1.0\columnwidth]{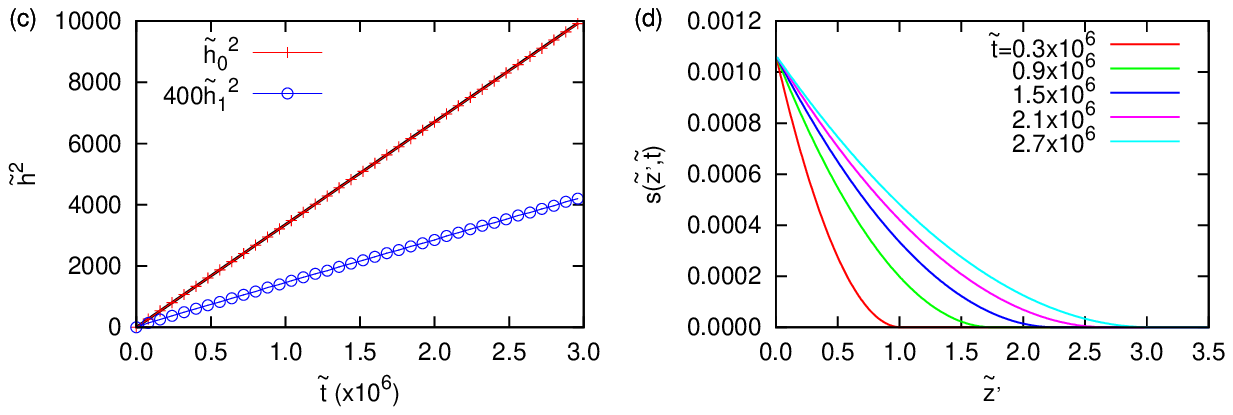}\\
\includegraphics[width=1.0\columnwidth]{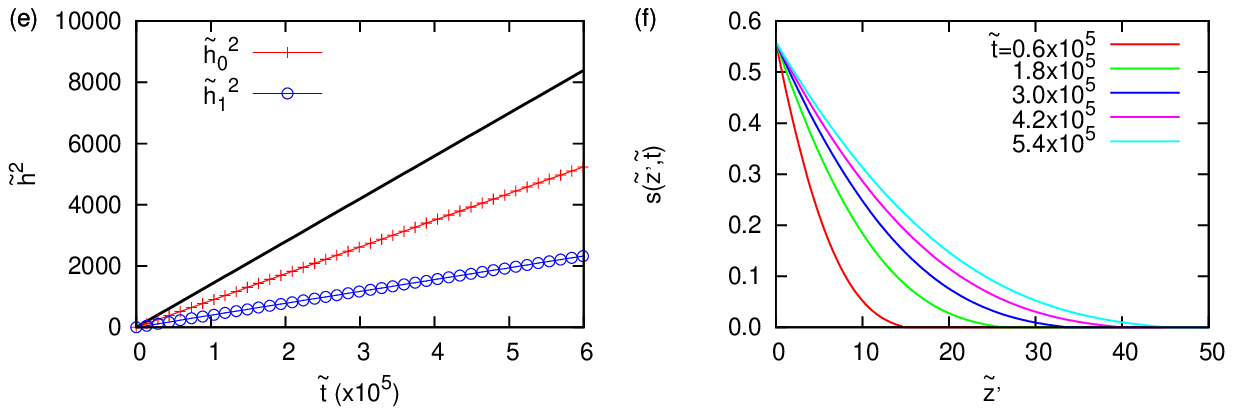}
\caption{The schematic pictures of the two different finger profile: case I (a) and case II (b). (c) The time evolution of the bulk length $h_0(t)$ and the finger length $h_1(t)$ for $\beta=0.01$ and $\alpha=10^{-5}$. In this case, the configuration is case I. (d) Time evolution of the finger profile for case I configuration. (e) Same as (c) but for case II configuration, $\beta=0.01$ and $\alpha=0.05$.  (f) Time evolution of the finger profile for case II configuration. \change{The black lines in (c) and (e) are the Lucas-Washburn predictions.}}
\label{fig:profile}
\end{figure}

We now apply the above theory for a nearly rectangular channel shown in Fig.~\ref{fig:profile}.
The channel has a cross section of thin hexagon consisting of two short sides of length $b$ and four long sides of length $a$ with very small aspect ratio($\beta = b/a \ll 1$).
The four corners of the channel have an angle $\pi/2+\alpha$ with $\alpha \ll 1$.

The equilibrium profile of saturation in such a channel was studied in the previous paper \cite{2021_rect_sstar}, according to which there are two possible configurations.
If $\alpha$ is zero (i.e.,the case of perfect rectangle) or very small, the bulk coexists with
four fingers as shown in Fig.~\ref{fig:profile}(a).
If $\alpha$ exceeds some critical value $\alpha_{\rm crit}$, which is a function of the aspect ratio $\beta$,  the bulk starts to coexist with two fingers  as shown in Fig.~\ref{fig:profile}(b).
We shall denote these two situations by case I and case II, respectively.


The dynamics of capillary filling is calculated by solving the dimensionless form of Eqs. (\ref{eq:diffusion}) and (\ref{eq:jstar}) numerically \cite{suppl} and the results are shown in  Fig.~\ref{fig:profile}. 
\change{The dimensionless parameters are transformed by $\tilde{h}_0 = h/a$, $\tilde{z}' = z'/a$, $\tilde{t} = t/( a \eta / \gamma)$.}

Figure~\ref{fig:profile}(c) and (d) show the results of the case I.
Here $\beta=0.01$ and $\alpha=10^{-5}$.
Figure~\ref{fig:profile}(c) shows the plot of $h_0^2(t)$ and $h_1^2(t)$ against time $t$.
(Here $400 h_1^2(t)$ is plotted against $t$ since $h_1^2(t)$ is too small to be seen.)
Both curves are straight, indicating that both the bulk flow and the finger flow
satisfy the Lucas-Washburn scaling of $t^{1/2}$.
\change{Note that when gravity is considered, the asymptotic behavior of the finger obeys a different $t^{1/3}$ scaling \cite{Tang1994, Higuera2008, Ponomarenko2011, 2020_onethird}.}
Figure \ref{fig:profile}(d) shows the time-evolution of the finger profile $s(z',t)$.
$s(z',t)$ starts from small value $s^*=0.00107$ at $z'=0$ and decrease to 0.
Since $s(z',t)$ is small in the entire finger region,  the effect of finger on the bulk flow
is small.
Similar conclusions have been obtained for the square tube\cite{2018_square}, where the correction to the Lucas-Washburn prefactor was about 5 \%.
In the present case, \change{the correction is much smaller}, and is negligible \cite{suppl}.

Figure~\ref{fig:profile}(e) and (f) show the dynamics of the case II.
Here $\beta=0.01$ and $\alpha=0.05$.
In this case, the finger part starts as two fingers at $z'=0$, which then split into
four fingers near the end of the fingers, i.e., near $z'=h_1$ as shown in Fig.~\ref{fig:profile}(b).
However, at the splitting point, $s$ and $f(s)$ are both continuous \cite{suppl}. Therefore no special consideration is needed for the splitting point.

\begin{figure}[htbp]
\includegraphics[width=0.8\columnwidth]{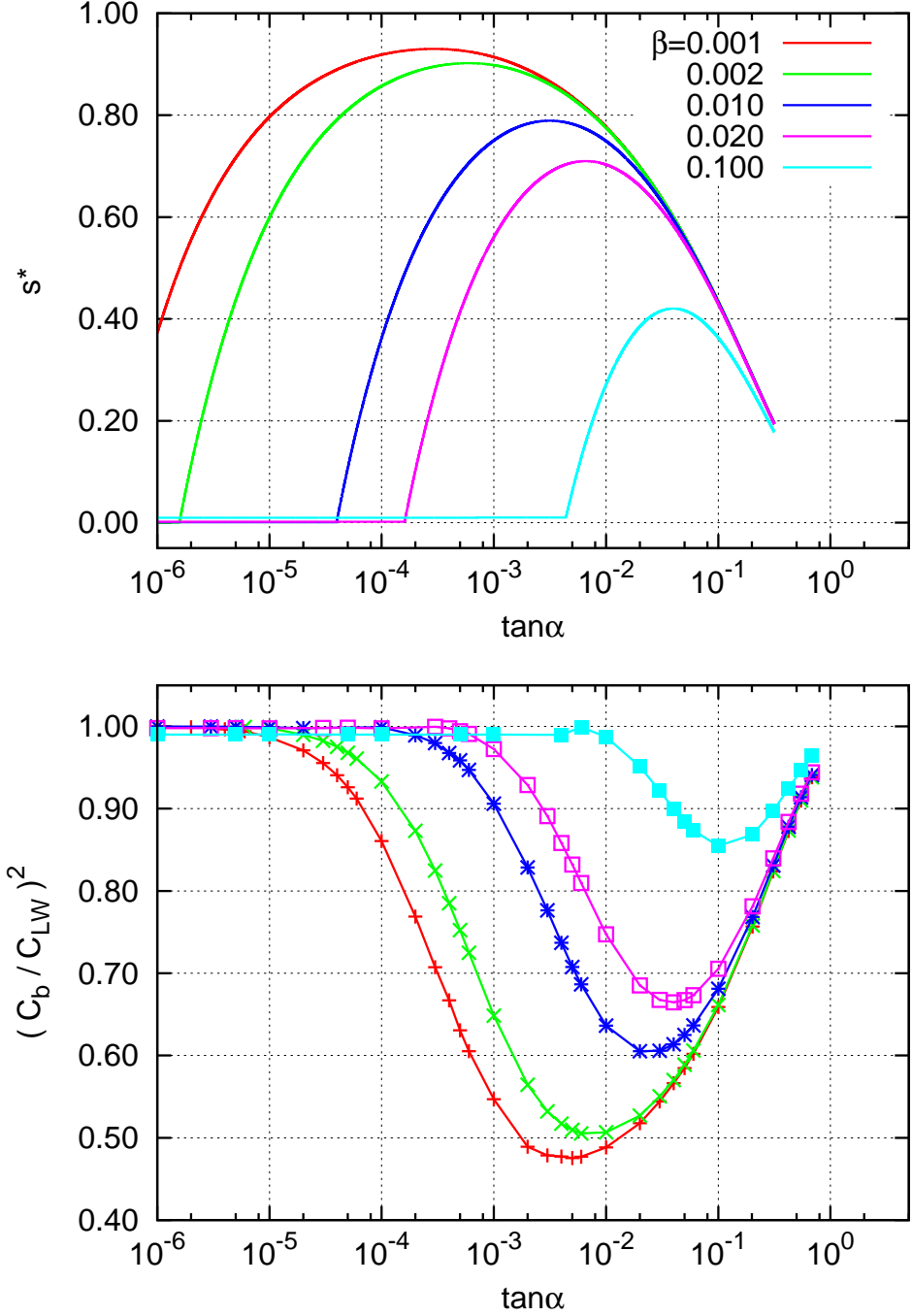}
\caption{(a) The value of $s^*$ as a function of the angle $\alpha$, for different aspect ratio $\beta$. (b) The reduction of the bulk factor $(\mathscr{C}_{\rm b}/\mathscr{C}_{\rm LW})^2$.}
\label{fig:sstar}
\end{figure}

Special attention is needed at the transition point from bulk to finger, where $s(z,t)$ changes discontinuously from 1 to $s^*$.
In the previous paper \cite{2021_rect_sstar}, we have shown that $s^*$ is very sensitive to the angle $\alpha$ and this is shown in Fig. \ref{fig:sstar}(a) which is a replot of the previous calculation.
It is seen that a small change of $\alpha$ causes a very large change of $s^*$.
Unlike the case I, which is shown by the flat part of the $s^*$ curve in the range of $\alpha < \alpha_{\rm crit}$, $s^*$ in case II is large, ($s^* \simeq 1$).
Such thick finger is expected to affect the flow of the bulk part significantly.

Figure~\ref{fig:profile}(e) shows the growth of the bulk part and the finger part in time.
Again both $h_0^2(t)$ and $h_1^2(t)$ increase linearly with time, indicating the Lucas-Washburn scaling $h_0(t) \sim t^{1/2}$, $h_1(t) \sim t^{1/2}$.
Unlike the case I, the prefactor of bulk flow in case II is much less than the prediction (\ref{eq:CLW}), indicating that thick fingers slow down the filling speed significantly.
The finger part advances ahead of the bulk part, utilizing portion of the interfacial
energy difference.
This gives smaller capillary driving force, and makes the bulk flow slower.
This reduction becomes prominent when the finger flow becomes significant, i.e., when $s^*$ is large.

To quantify the effect of finger part on the filling speed, we performed calculations varying the aspect-ratio $\beta$ and the angle $\alpha$.
By fitting the bulk length and the finger length with $h_0(t) = \mathscr{C}_{\rm b} t^{1/2}$ and $h_1(t) = \mathscr{C}_{\rm f} t^{1/2}$, we can obtain the bulk prefactor $\mathscr{C}_{\rm b}$ and finger prefactor $\mathscr{C}_{\rm f}$.
The results for the dynamics are shown in Fig.~\ref{fig:sstar}(b), where the reduction $(\mathscr{C}_{\rm b} / \mathscr{C}_{\rm LW} )^2$ is plotted as a function of $\tan\alpha$ for various values of $\beta$.
The nearly flat lines at small $\tan\alpha$ correspond to case I (four fingers).
Only small reduction in the prefactor is observed in this case.
The curved lines with a minimum correspond to case II (two fingers).
Small change of the $\tan\alpha$ value leads to a dramatic change of the value of $s^*$, and the finger flow becomes significant.
As a consequence, the bulk flow is reduced with respect to the Lucas-Washburn prediction (\ref{eq:CLW}).



\begin{figure}[htp]
\includegraphics[width=1.0\columnwidth]{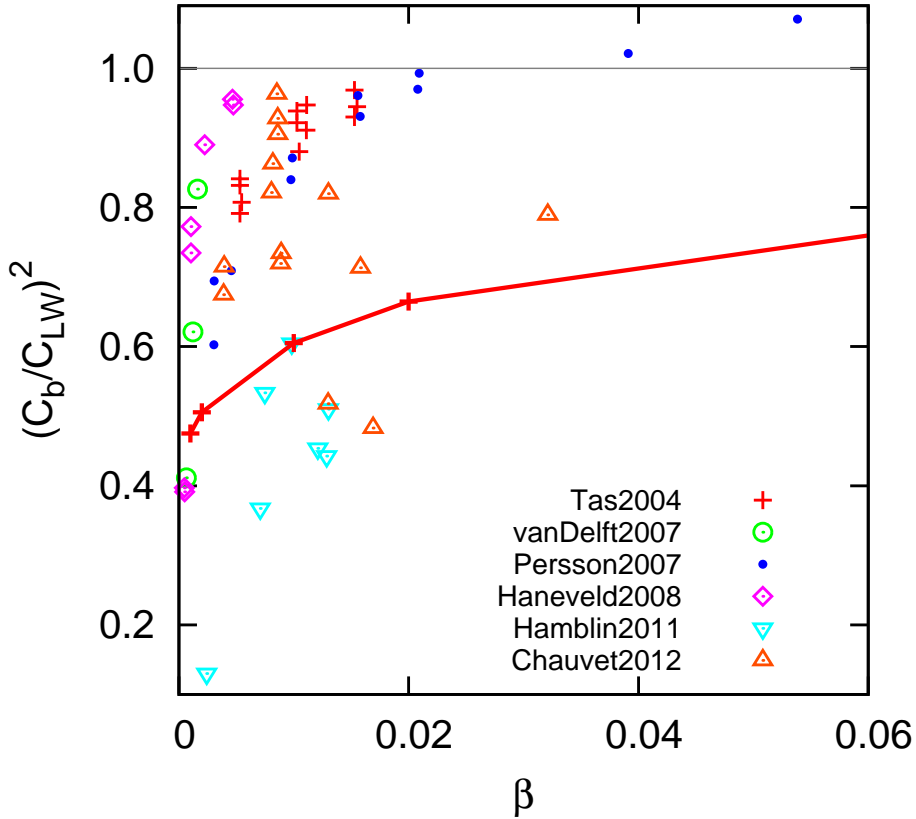}
\caption{Comparison to the experimental measurement. The $x$-axis is the aspect ratio $\beta$: small $\beta$ corresponds to an elongated rectangular cross-section. The $y$-axis is $(\mathscr{C}_{\rm b}/\mathscr{C}_{\rm LW})^2$, the reduction in the prefactor with respect to Lucas-Washburn prediction (\ref{eq:CLW}).}
\label{fig:exp}
\end{figure}

We now compare our results with the experimental data in literature \cite{Tas2004, vanDelft2007, Persson2007, Haneveld2008, Hamblin2011, Chauvet2012}.
Figure \ref{fig:exp} shows the experimentally obtained prefactors $(\mathscr{C}_{\rm b} / \mathscr{C}_{\rm LW})^2$ plotted against the aspect ratio $\beta$.
It is seen that there is a significant scattering for the reported prefactors $\mathscr{C}_{\rm b}$, but most of them are smaller than the theoretical value $\mathscr{C}_{\rm LW}$ which accounts for the bulk part only.
This is consistent with our theoretical predictions that (1) the imbibition speed is slowed down by the finger part which goes ahead of the bulk part, and that (2) for rectangular channel, the effect of the slowing down becomes very sensitive to tiny deviation of the tube shape.
The red line in Figure \ref{fig:exp} shows the minimum value of $(\mathscr{C}_{\rm b} / \mathscr{C}_{\rm LW})^2$ of our theoretical calculation when $\alpha$ is varied.
It is seen that most experimental data are between the red-line and the gray-line of $(\mathscr{C}_{\rm b} / \mathscr{C}_{\rm LW})^2=1$.
This indicates that the deviation from the classical Lucas-Washburn theory is perhaps due to the small deviation of the channel shape from perfect rectangles.

\section{Conclusion}
To summarize, we develop a general framework to analyze the coupling between the finger flow and the bulk flow in channels with corners.
We then apply this framework to study the capillary filling in a nearly rectangular channel, which was used for most capillary experiments at nanometer scale.
Our numerical results indicate that both the finger flow and the bulk flow satisfy the Lucas-Washburn scaling law, as their lengths increase with time as $t^{1/2}$.
However, the prefactor can be different than the Lucas-Washburn prediction:
The prefactor for the bulk flow is reduced considerably when the finger flow becomes significant.
Our results offer an explanation for the experimental observation that the prefactor is smaller than the theoretical prediction without considering the finger flow.


\paragraph*{Acknowledgments.}
This work was supported by the National Natural Science Foundation of China (NSFC) through the Grant No. 21774004 (to J.Z.).
M.D. acknowledges the financial support of the Chinese Central Government in the Thousand Talents Program.

\bibliography{wetting}

\end{document}


\title{Supporting Information: \\
\emph{Wetting dynamics in an angular channel}}

\author{Chen Zhao}
\affiliation{Center of Soft Matter Physics and its Applications, Beihang University, Beijing 100191, China}
\affiliation{School of Physics, Beihang University, Beijing 100191, China}

\author{Tian Yu}
\affiliation{State Key Laboratory of Hydroscience and Engineering, Tsinghua University,
Beijing 100084, China}

\author{Jiajia Zhou}
\email[]{zhouj2@scut.edu.cn}
\affiliation{South China Advanced Institute for Soft Matter Science and Technology, South China University of Technology, Guangzhou 510640, China}
\affiliation{School of Molecular Science and Engineering, Guangzhou International Campus, South China University of Technology, Guangzhou 511442, China}

\author{Masao Doi}
\affiliation{Center of Soft Matter Physics and its Applications, Beihang University, Beijing 100191, China}


\maketitle

\section{Square tube}

In this section, we calculate the dynamics of the capillary filling in a squared tube.
We have presented the results in Ref. \cite{2018_square} by solving the coupled PDE/ODE.
Here we discuss the bulk/finger coupling using the notation developed in the main text.

For a square tube with the side length $2a$, we have
\begin{eqnarray}
  s^* &=& (1-\sqrt{\pi/4})/(1+\sqrt{\pi/4}) \simeq 0.0603178 \\
  f(1) &=& -8 a \gamma = - A_0 a\gamma,
           \quad A_0 = 8 \\
  f(s) &=& -8 \sqrt{1-\frac{\pi}{4}} s^{1/2} a\gamma = - A_1 s^{1/2} a\gamma,
           \quad A_1 = 8 \sqrt{1-\frac{\pi}{4}} \\
  \label{eq:zeta1_square}
  \zeta(1) &=& B_0 \eta,
           \quad B_0 \simeq 28.4558 \\
  \zeta(s) &=& B_1 s^{-2} \eta,
           \quad B_1 \simeq 78.5031 \\
  D(s) &=& \frac{f''(s)}{\zeta(s)} = \frac{A_1}{4B_1} s^{1/2} \frac{a\gamma}{\eta}
\end{eqnarray}
Here the free energy quantities are calculated by geometric consideration \cite{2018_square}.
The frictional coefficients are calculated using the method in Refs.~\cite{Ransohoff1988, 2018_square}.
Note we have used a slightly different notation.
In Ref.~\cite{2018_square}, the friction coefficient $\xi(s)$ is coupled to the volume flux $Q$, and the dissipation function is written as
\begin{equation}
  \Phi = \int \ud z \, \frac{1}{2} \xi Q^2.
\end{equation}
In the main text, we use the friction coefficient $\zeta(s)$ coupled to the cross-section-averaged flux $j=Q/S_0$, where $S_0$ is the cross-section area of the tube/channel.
The dissipation function written in term of $\zeta(s)$ is
\begin{equation}
  \Phi = \int \ud z \, \frac{1}{2} \zeta j^2.
\end{equation}
These two coefficient are related by
\begin{equation}
  \xi(s) = \zeta(s) S_0^2 .
\end{equation}

\subsection{Time-evolution equations}

We perform the following transformation to make the PDE/ODE dimensionless
\begin{eqnarray}
  t &=& \bar{t} \frac{a \eta}{\gamma}, \\
  z' &=& \bar{z}' a , \\
  j &=& \bar{j} \frac{\gamma}{\eta} ,\\
  D &=& \bar{D} \frac{a\gamma}{\eta} .
\end{eqnarray}
The dimensionless PDE is
\begin{equation}
  \label{eq:pde_square}
  \frac{\partial s}{\partial \bar{t}} = \frac{\partial}{\partial \bar{z}'}
    \left[ \frac{A_1}{4B_1} s^{1/2} \frac{\partial s}{\partial \bar{z}'} \right]
    + \dot{\bar{h}}_0 \frac{\partial s}{\partial \bar{z}'}
\end{equation}
For the ODE, we need the following quantities
\begin{eqnarray}
  D(s^*) &=& \frac{A_1}{4B_1} (s^*)^{1/2} \frac{a\gamma}{\eta} \\
  - \frac{f'(s^*)}{\zeta(1)} &=& \frac{A_1}{2B_0} (s^*)^{-1/2} \frac{a\gamma}{\eta}.
\end{eqnarray}
The dimensionless ODE is
\begin{equation}
  \label{eq:ode_square}
  \dot{\bar{h}}_0 = \frac{1}{1-s^*} \left[ \frac{A_1}{4B_1} (s^*)^{1/2} \frac{\partial s}{\partial \bar{z}'} \Big|_{\bar{z}'=0} + \frac{A_1}{2B_0} (s^*)^{-1/2} \frac{1}{\bar{h}_0} \right] .
\end{equation}
The coupled equations (\ref{eq:pde_square}) and (\ref{eq:ode_square}) agree with the results in Ref.~\cite{2018_square}.

\section{Perfect rectangular channel}

For a rectangular channel with sides $2a$ and $b$ ($2a>b$), we define an aspect ratio
\begin{equation}
  \beta = \frac{b}{a}.
\end{equation}
The square tube discussed in the last section corresponds to the special case of $\beta=2$.
We shall express all quantities in term of $\beta$.
\begin{eqnarray}
  s^* &=& \frac{ \left[ (2+\beta) - \sqrt{ (2+\beta)^2 - 8 \beta(1-\frac{\pi}{4}) } \right]^2 }
          { 8 \beta ( 1- \frac{\pi}{4} )} \\
  f(1) &=& - (4a+2b) \gamma = - (4+2\beta) a \gamma = - A_0 a\gamma,
           \quad A_0 = 4 + 2\beta \\
  f(s) &=& - \sqrt{ 32 \beta (1-\frac{\pi}{4}) } s^{1/2} a\gamma = - A_1 s^{1/2} a\gamma,
           \quad A_1 = \sqrt{32 \beta (1-\frac{\pi}{4})}
\end{eqnarray}

\begin{figure}[htp]
\includegraphics[width=0.7\textwidth]{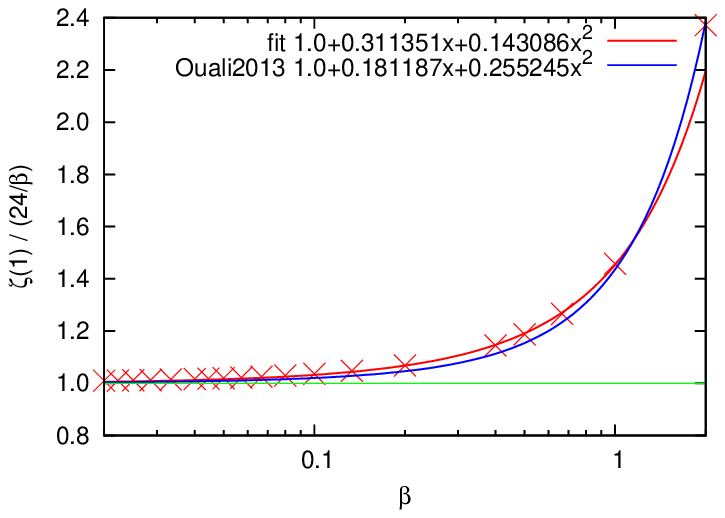}
\caption{Numerical results of function $\zeta_c^{-1}(\beta)$ and different fits.}
\label{fig:friction_rect}
\end{figure}

The dissipation function is more complicated.
In the case of large aspect ratio $\beta$, i.e., $2a \gg b$, one can neglect the edge effect (due to the two short $b$-sides) and use the lubrication theory to obtain
\begin{equation}
  \zeta(1) = \frac{24}{\beta} \eta.
\end{equation}
This becomes less accurate when $\beta$ approaches $1$.
In Ref.~\cite{Ouali2013}, Ouali \emph{et al.} derived a Fourier series solutions and also presented a simple interpolation formula (their $\epsilon$ is equal to our $\beta/2$)
\begin{equation}
  \zeta(1) = \frac{24}{\beta \zeta_c(\beta)} \eta,
  \quad \zeta_c^{-1}(\beta) = 1 + 0.181187 \beta + 0.255245 \beta^2 .
\end{equation}
They stated the agreement is good to around 3\% when $\beta < 1.2$ (or $\epsilon < 0.6$).
We performed Matlab calculation and the results are shown in Fig.~\ref{fig:friction_rect}.
For $\beta<1.0$, the following fit has a better agreement
\begin{equation}
  \zeta(1) = \frac{24}{\beta \zeta_c(\beta)} \eta,
  \quad \zeta_c^{-1}(\beta) = 1 + 0.311351 \beta + 0.143086 \beta^2 .
\end{equation}
This is what we shall use in the following calculation
\begin{equation}
  \zeta(1) = B_0 \eta, \quad B_0 = \frac{24}{\beta} ( 1 + 0.311351 \beta + 0.143086 \beta^2).
\end{equation}
Note this expression gives a different value for $\beta=2$ than Eq.~(\ref{eq:zeta1_square}) because the fitting is done for $\beta<1.0$.

For bulk-only case, the Lucas-Washburn factor is
\begin{eqnarray}
  \mathscr{C}^2_{\rm LW} &=& \frac{2 |f(1)|}{\zeta(1)}
  = \frac{ 4+2\beta }{12 \zeta_c(\beta)} \frac{b \gamma}{\eta} \\
  &\simeq&  \frac{b \gamma}{ 3 \eta} \quad (\beta \rightarrow 0)
\end{eqnarray}
The $\beta\rightarrow 0$ expression is the theoretical result used in many experimental reports.

For partial saturation, the friction coefficient is
\begin{equation}
  \zeta(s) = B_1 s^{-2} \eta, \quad B_1 \simeq 78.5031
\end{equation}
which is the same as the square tube.

\begin{figure}[htp]
\includegraphics[width=1.0\columnwidth]{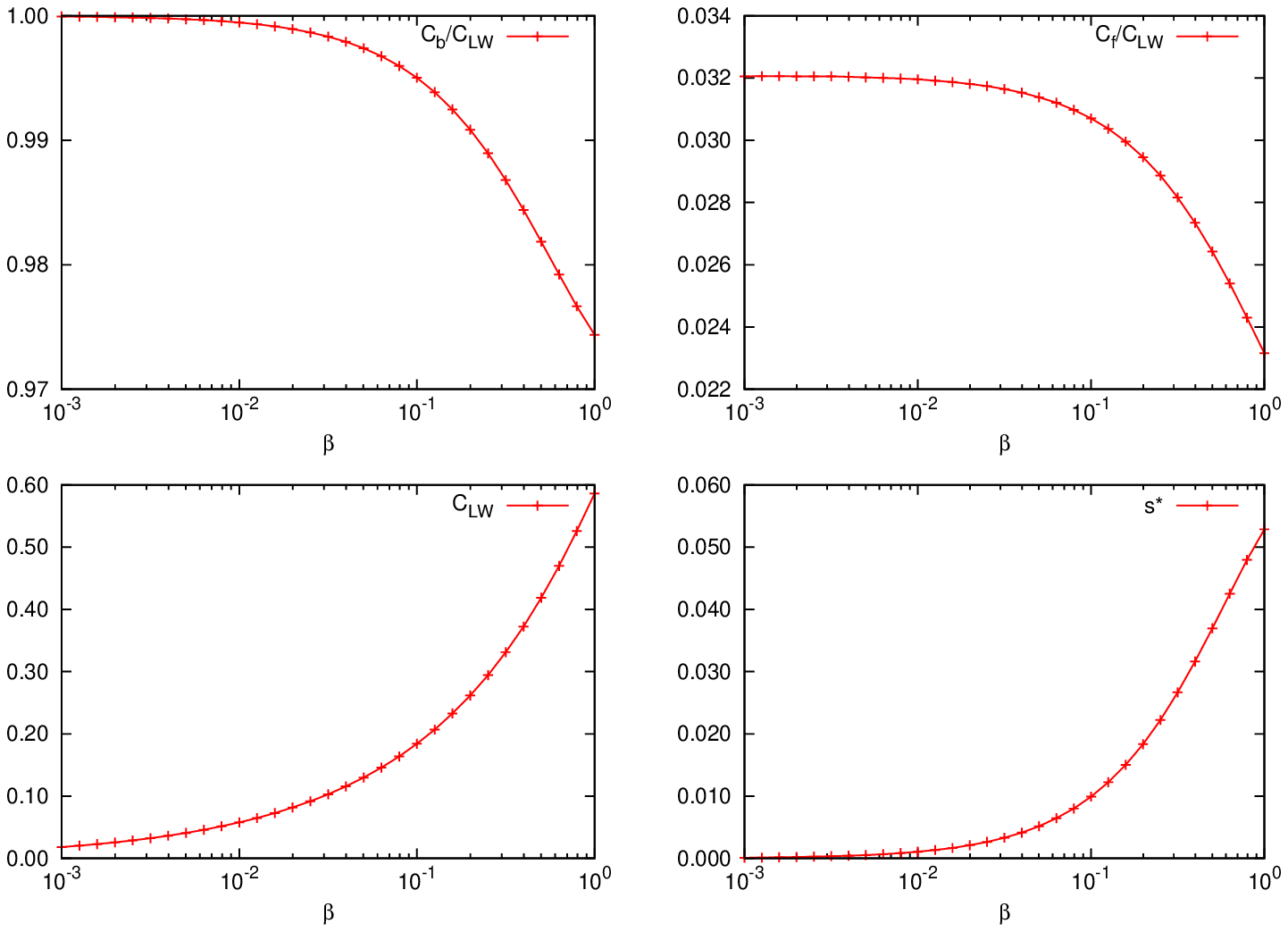}
\caption{The Lucas-Washburn factors for a perfect rectangular channel.}
\label{fig:LW_rect}
\end{figure}

Figure \ref{fig:LW_rect} shows the numerical results.
For perfect rectangular channels, the bulk factor $\mathscr{C}_{\rm b}$ is only slightly reduced with respect to the Lucas-Washburn value $\mathscr{C}_{\rm LW}$.

\section{Nearly rectangular channel}

Here we discuss the nearly rectangular channel, which the four corners have an angle slightly larger than the normal angle, $\pi/2+\alpha$.

\subsection{case I}

When $\alpha$ is small, there are four separated fingers along each \change{corner} [Fig. 2(a) in the main text).
The $s^*$ value is given by
\begin{equation}
  s^* = \frac{ (\tan\alpha + \sec\alpha)  \left( (2+\beta) - \sqrt{ (2+\beta)^2 - \frac{8\cos\alpha (\sin\alpha + \beta)}{( \tan\alpha + \sec\alpha)} \left[ 1 - (\frac{\pi}{4}-\frac{\alpha}{2})(\tan\alpha + \sec\alpha) \right] } \right)^2 }{ 8 \cos\alpha (\sin\alpha + \beta)
      \left[ 1 - (\frac{\pi}{4}-\frac{\alpha}{2})(\tan\alpha + \sec\alpha) \right] } .
\end{equation}

The following quantities are required for the calculation:
\begin{eqnarray}
  f(1) &=& - ( 4 + 2\beta ) a\gamma = - A_0 a \gamma, \quad A_0 = 4 + 2\beta \\
  \label{eq:f_case1}
  f(s) &=& - A_1 s^{1/2} a \gamma, \quad A_1 = \left[ \frac{32 \cos\alpha ( \sin\alpha + \beta)}
           {(\tan\alpha + \sec\alpha)} [ 1-(\frac{\pi}{4} - \frac{\alpha}{2})
           (\tan\alpha + \sec\alpha) ] \right]^{1/2} \\
  \label{eq:zeta1_case1}
  \zeta(1) &=& B_0 \eta, \quad B_0 = \frac{12}{\tan\alpha} \ln ( 1 + \frac{2\sin\alpha}{\beta}) \\
  \label{eq:zeta_case1}
  \zeta(s) &=& B_1 s^{-2} \eta, \quad B_1 = 78.4201 + 9.93084 \alpha + 19.84 \alpha^2 .
\end{eqnarray}

The parameter $B_0$ is calculated using the lubrication approximation, which we will explain \change{later}.
The parameter $B_1$ is computed using the method in Ref.~\cite{Ransohoff1988}, and this parameter only depends on the angle $\alpha$ and does not depend on the aspect ratio $\beta$.

\subsection{case II}

In this case, the fingers are composed of two fingers near $s^*$, then become four fingers near the tip [Fig. 2(b) in the main text].

The $s^*$ is given by \cite{2021_rect_sstar}
\begin{equation}
  s^* = \frac{ ( \sin\alpha + \frac{1}{2}\beta )^2
      \left[ -1 + \frac{ 2( 1 - \sqrt{ 1 - \cos^2\alpha [ 1 - (\frac{\pi}{2} - \alpha) \tan\alpha ] } )}
      { \cos^2 \alpha [ 1 - (\frac{\pi}{2} - \alpha) \tan\alpha ] } \right] - \frac{1}{4} \beta^2 }
      { \sin\alpha ( \sin\alpha + \beta )} .
      \label{eq:sstar_case2}
\end{equation}

We also need two quantities $f(s)$ and $\zeta(s)$.
Here we need to know an intermediate quantity $\ell$, which is the length on $a$-side covered by the fluid.
The saturation $s$ is related to $\ell$ by
\begin{eqnarray}
  s &=& \frac{ \tan\alpha \left[ 1 - (\frac{\pi}{2} - \alpha) \tan\alpha \right]}
                   { \cos\alpha (\sin\alpha + \beta) } \Big( \frac{\ell}{a} \Big)^2 \nonumber \\
   && + \frac{ {\sec\alpha} \left[ 1 - (\frac{\pi}{2} - \alpha) \tan\alpha \right]}
                   { \cos\alpha (\sin\alpha + \beta) } \beta \Big( \frac{\ell}{a} \Big)
      + \frac{ \tan\alpha - (\frac{\pi}{2} - \alpha) {\sec^2\alpha} }
              { 4 \cos\alpha ( \sin\alpha + \beta)} \beta^2 .
      \label{eq:s_case2}
\end{eqnarray}
Once $\ell$ is obtained by solving the above equation, the free energy density and the friction coefficient are
\begin{eqnarray}
  \label{eq:f_case2}
  f(s) &=& \left[ - 4 \left[ 1 - (\frac{\pi}{2} - \alpha) \tan\alpha \right] \frac{\ell}{a}
                  - 2 \left[ 1 - (\frac{\pi}{2} - \alpha) {\sec\alpha} \right] \beta \right] a \gamma , \\
  \label{eq:zeta_case2}
  \zeta(s) &=& \frac{12\eta}{s^2 \tan\alpha} \ln \left(
      1 + \frac{2\ell \sin\alpha}{b} \right).
\end{eqnarray}

\begin{figure}[htp]
\includegraphics[width=0.6\columnwidth]{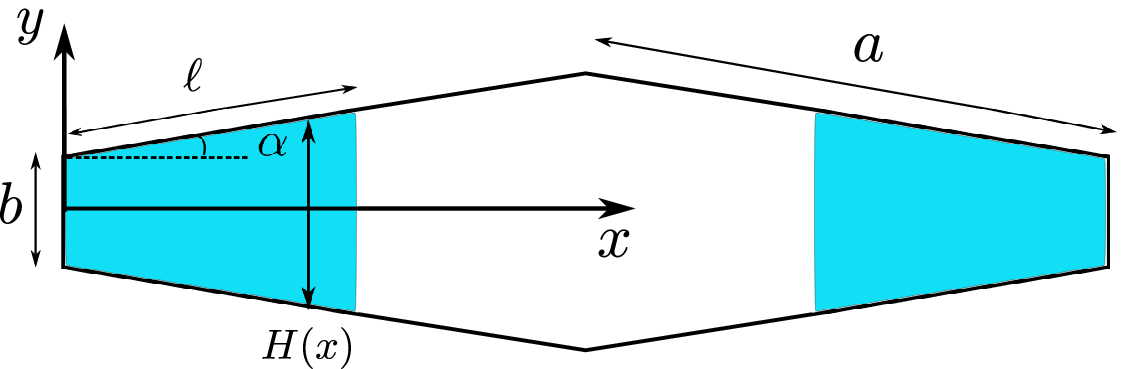}
\caption{Lubrication calculation of the friction coefficient $\zeta(s)$.}
\label{fig:lub}
\end{figure}

The friction coefficient (\ref{eq:zeta_case2}) is calculated using lubrication theory.
For the setup shown in Fig.~\ref{fig:lub}, the thickness is given by
\begin{equation}
  H(x) = b + 2 x \tan\alpha.
\end{equation}
The dissipation function is then given by
\begin{eqnarray}
  \Phi &=& \frac{1}{2} 2 \int \ud z \int_0^{\ell \cos\alpha} \ud x  \frac{12 \eta}{H(x)} v_z^2 \nonumber \\
  &=& \frac{1}{2} \int \ud z \frac{12\eta}{\tan\alpha} \ln \left(
      1 + \frac{2\ell \sin\alpha}{b} \right) v_z^2 .
\end{eqnarray}
The average velocity $v_z = j/s$, thus
\begin{eqnarray}
  \Phi &=& \frac{1}{2} \int \ud z \frac{12\eta}{s^2 \tan\alpha} \ln \left(
      1 + \frac{2\ell \sin\alpha}{b} \right) j^2 \\
  \zeta(s) &=& \frac{12\eta}{s^2 \tan\alpha} \ln \left(
      1 + \frac{2\ell \sin\alpha}{b} \right).
\end{eqnarray}

Setting $s=1$, $\ell=a$, we get the expression (\ref{eq:zeta1_case1}) for $\zeta(1)$.

The saturation of the finger is a monotonic decreasing function from $s^*$ to the finger tip $s=0$.
When $s<s_{\rm c1}$, the two-finger configuration is not possible and there are four fingers near the tip.
The value of $s_{c1}$ is given by
\begin{equation}
  s_{\rm c1} = \frac{(\tan\alpha + \sec\alpha)}{2 \cos\alpha ( \sin\alpha + \beta)} \left[ 1 - (\frac{\pi}{4}-\frac{\alpha}{2}) (\tan\alpha + \sec\alpha) \right] \beta^2 .
\end{equation}

When $s<s_{\rm c1}$, the free energy density and the frictional coefficient are given by Eqs. (\ref{eq:f_case1}) and (\ref{eq:zeta_case1}), respectively.

\bibliography{wetting}